\documentclass[mathleft
]{an}
\usepackage{graphicx}
\usepackage{times}
\begin{document}

\Pagespan{789}{}
\Yearpublication{2010}%
\Yearsubmission{2010}%
\Month{11}%
\Volume{999}%
\Issue{88}%

\title{Enhancing the signal-to-noise ratio of solar-like targets}

\author{R.A. Garc\'\i a\inst{1}\fnmsep\thanks{Corresponding author:
  \email{rgarcia@cea.fr}\newline}
\and S. Mathur\inst{2}
\and J. Ballot\inst{3}
\and C. R\'egulo\inst{4,5}
}
\titlerunning{Enhancing the signal-to-noise ratio of solar-like targets}
\authorrunning{Garc\'\i a, Ballot, Mathur \& R\'egulo}
\institute{
Laboratoire AIM, CEA/DSM-CNRS, Universit\'e Paris 7 Diderot, IRFU/SAp-SEDI, Centre de Saclay, 91191, Gif-sur-Yvette, France
\and 
High Altitude Observatory, NCAR, P.O. Box 3000, Boulder, CO 80307, USA
\and 
Laboratoire d'Astrophysique de Toulouse-Tarbes, Universit\'e de Toulouse, CNRS, F-31400, Toulouse, France
\and 
 Universidad de La Laguna, Dpto de Astrof\'isica, 38206, Tenerife, Spain
\and 
 Instituto de Astrof\'\i sica de Canarias, 38205, La Laguna, Tenerife, Spain
}

\received{}
\accepted{}
\publonline{later}


\abstract{The analysis of the first solar-like targets done by CoRoT has shown that the oscillation amplitudes are about 25$\%$ below the theoretical amplitudes while the convective backgrounds are up to three times higher than in the solar case (Michel et al. 2008). In such conditions, the comb-like structure of the acoustic modes has smaller signal-to-noise ratios than initially expected complicating the characterization of individual modes. In the present work we apply the curvelet filtering to the solar-like targets already observed by CoRoT as well as a partial reconstruction of the signal from the obtained spacing of the comb-like structure of the acoustic modes. It enables us to enhance the signal-to-noise ratio of the ridges in the Echelle diagrams. Finally, we study how the analysis of the p modes can be improved.
}

\maketitle

\section{Introduction}
The signal-to-noise ratio (SNR) of solar-like stars can be enhanced by taking advantage of the asymptotic properties of their acoustic  (p)  modes. In the Fourier spectrum of these stars, the peaks that correspond to the acoustic modes are almost equally spaced in a given frequency range. Here we describe two methods: the partial reconstruction of the signal and the Curvelet filtering and we apply them using  asteroseismic observations obtained by CoRoT (Convection, Rotation and planetary Transits). 

\section{Partial reconstruction of the signal}
To increase the SNR of the periodic structures (the modes) we perform a selective filtering of the power spectrum of the power spectrum (PSPS) (see for details: R\'egulo and Roca Cort\'es 2002; Mathur et al. 2010a).

\begin{figure}[!htbp]
\includegraphics[width=6cm, angle=90]{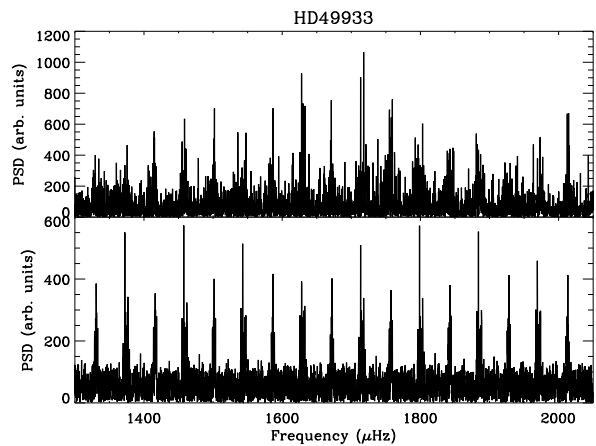}
\caption{Raw (top) and recovered (bottom) power spectrum density (PSD) of HD49933. The comb-like structure of the modes is enhanced in the recovered PSD.\label{HD49933}}
\end{figure}

We start by computing the PSPS of a short range in frequency ($< 1000\mu$Hz) around the maximum power of the p-mode band (the asymptotic part).  The resultant PSPS is filtered by multiplying it by a window function, which is equal to 1 for all the equally spaced bins around multiples of the large separation of the p modes (that should have been estimated before) starting at zero, while the rest of the bins are settled to 0. The inverse Fourier Transform of this filtered PSPS produces a ``recovered'' power spectrum of the stellar p modes with a higher SNR compared to the original one. We have applied this technique to two high SNR CoRoT targets: HD181420 (Barban et al. 2009) and HD49933 (Appourchaux et al. 2008; Benomar et al. 2009). An example of the results for the second star is shown in Fig.~\ref{HD49933}. 

\begin{figure*}[!htbp]
\includegraphics[width=6cm, angle=90]{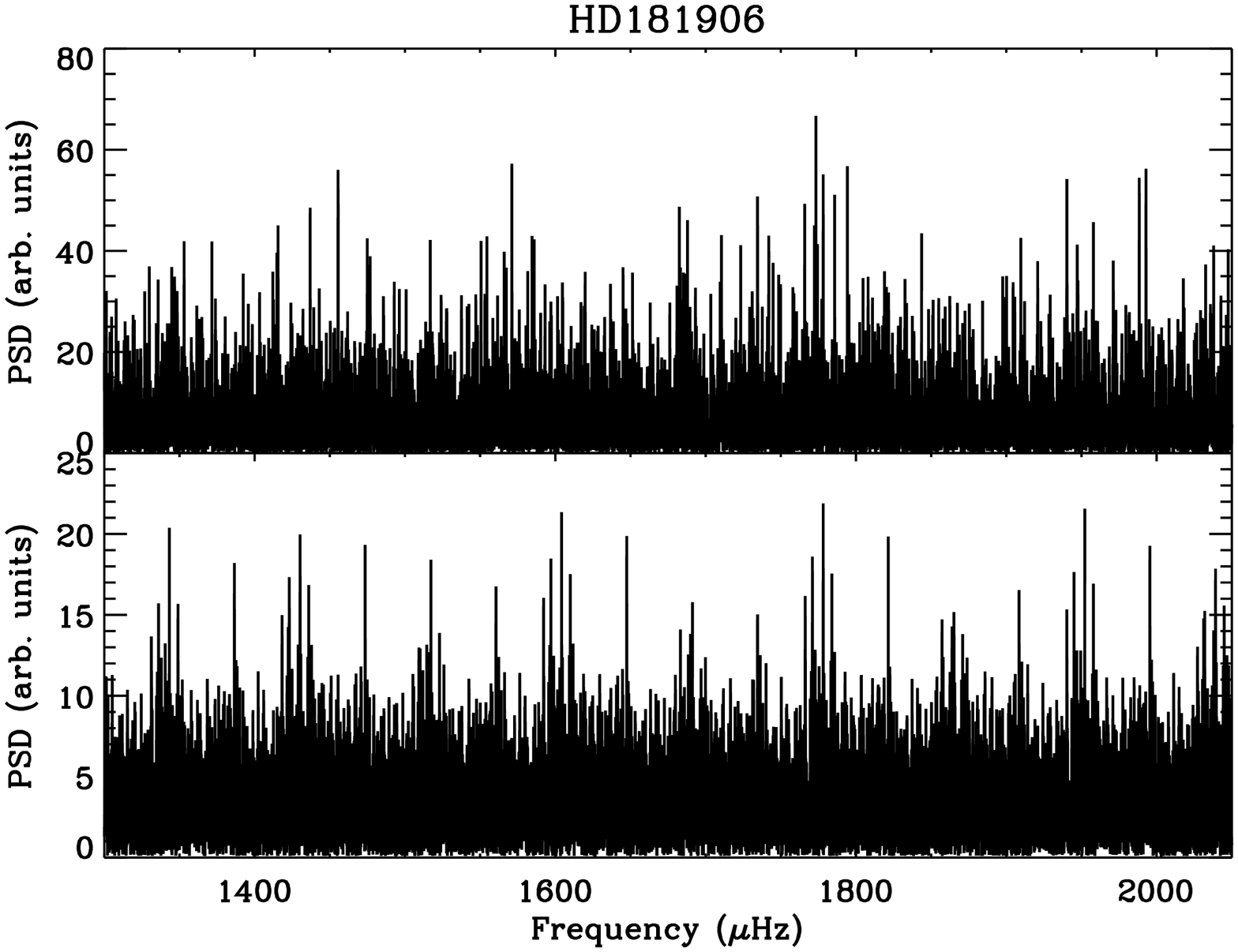}
\includegraphics[width=6cm, angle=90]{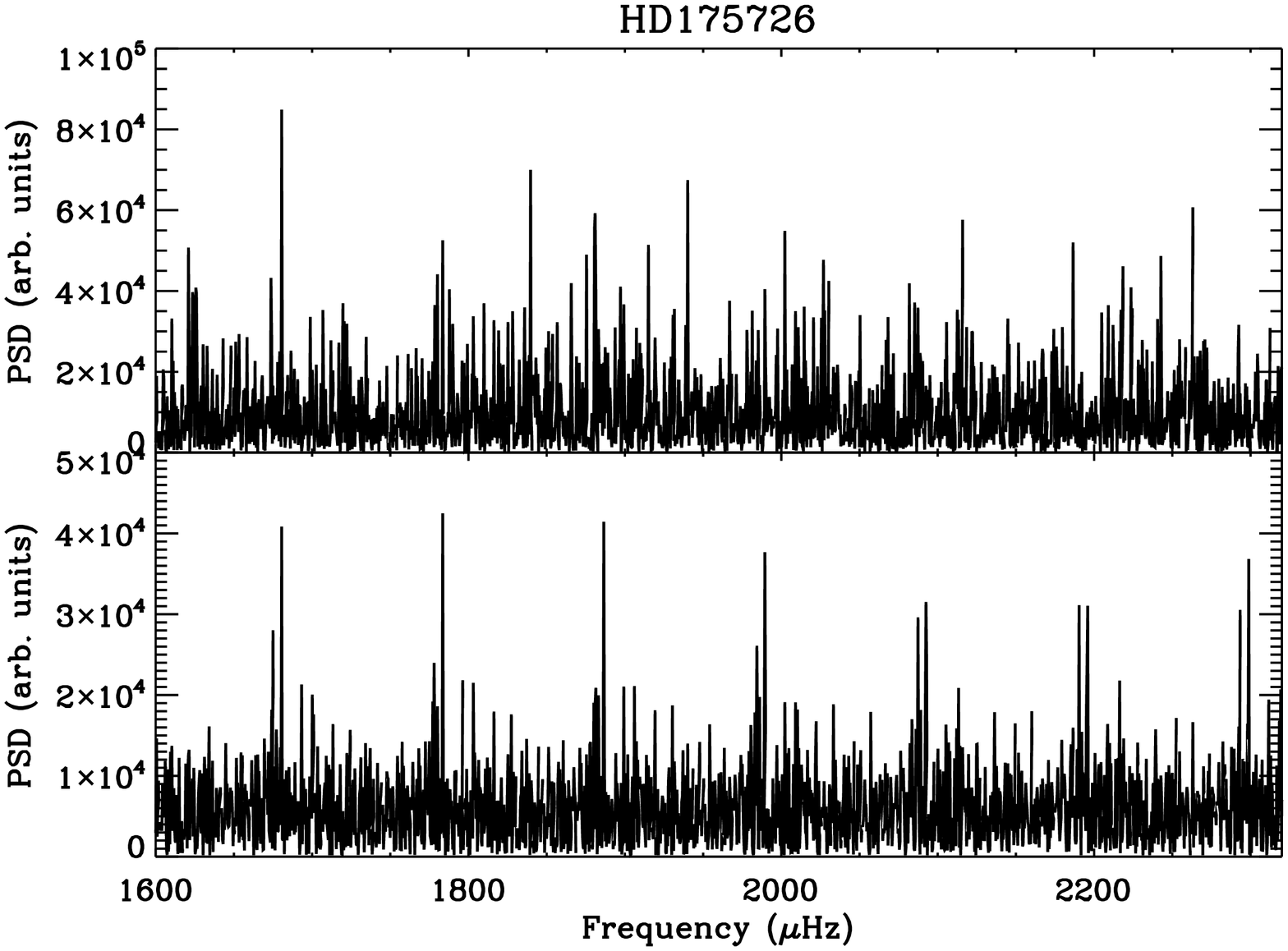}
\caption{Raw (top) and recovered (bottom) PSD of the low SNR targets HD181906 and HD175726.\label{fig2}}
\end{figure*}

Fig.~\ref{fig2} shows the raw and the recovered PSD of two other low SNR CoRoT targets: HD181906 (Garc\'\i a et al. 2009) and HD175726 (Mosser et al. 2009). In both cases, the modes are very difficult to be distinguished in the raw PSD but in the recovered ones, we can easily see the comb-like structure of the p modes.

This method tends to rigidify the modes (constant spacing) towards the borders of the considered region where the modes could be less asymptotic and the large separations could change significantly with frequency. The ÒrecoveredÓ amplitudes and line-widths of the modes are unfortunately modified, thus this method is only reliable, for the moment, to study the frequency distribution of p modes.

\section{Curvelet filtering}
It is commonly accepted that Echelle Diagrams can help in the process of mode tagging as it has already been done in early stages of helioseismology (e.g. Grec et al. 1981). To improve the SNR, Bedding et al. (2004) proposed to use this Echelle diagram by smoothing it in the vertical direction. However, this method only works well in the frequency region in which the ridglets are quasi-vertical, i.e., in the asymptotical regime. Even more, a very good a priori of the large separation is needed.

We propose here another denoising technique based on the use of {\it Curvelets}. Curvelet transforms are built to deal with curved structures of finite size in the image. A curvelet transform  (Cand\`es \& Donoho 1999) is a ridgelet transform (Cand\`es 1998) but used in a a localized manner: at a small scale any curved line can be approached by a straight line. Using standard wavelets,  many coefficients are needed to well define the curved figures while with the curvelets, a few coefficients are enough. 

The curvelet transform has been successfully tested in asteroseismology by an extensive use of  Monte-Carlo simulations and real solar data (Lambert et al. 2006). The filtering becomes more efficient when the SNR is small and it only needs a rough estimation of the large separation.

\begin{figure}[!htbp]
\includegraphics[width=6.3cm, trim = 1cm 3.cm 2cm 9cm, angle=-90]{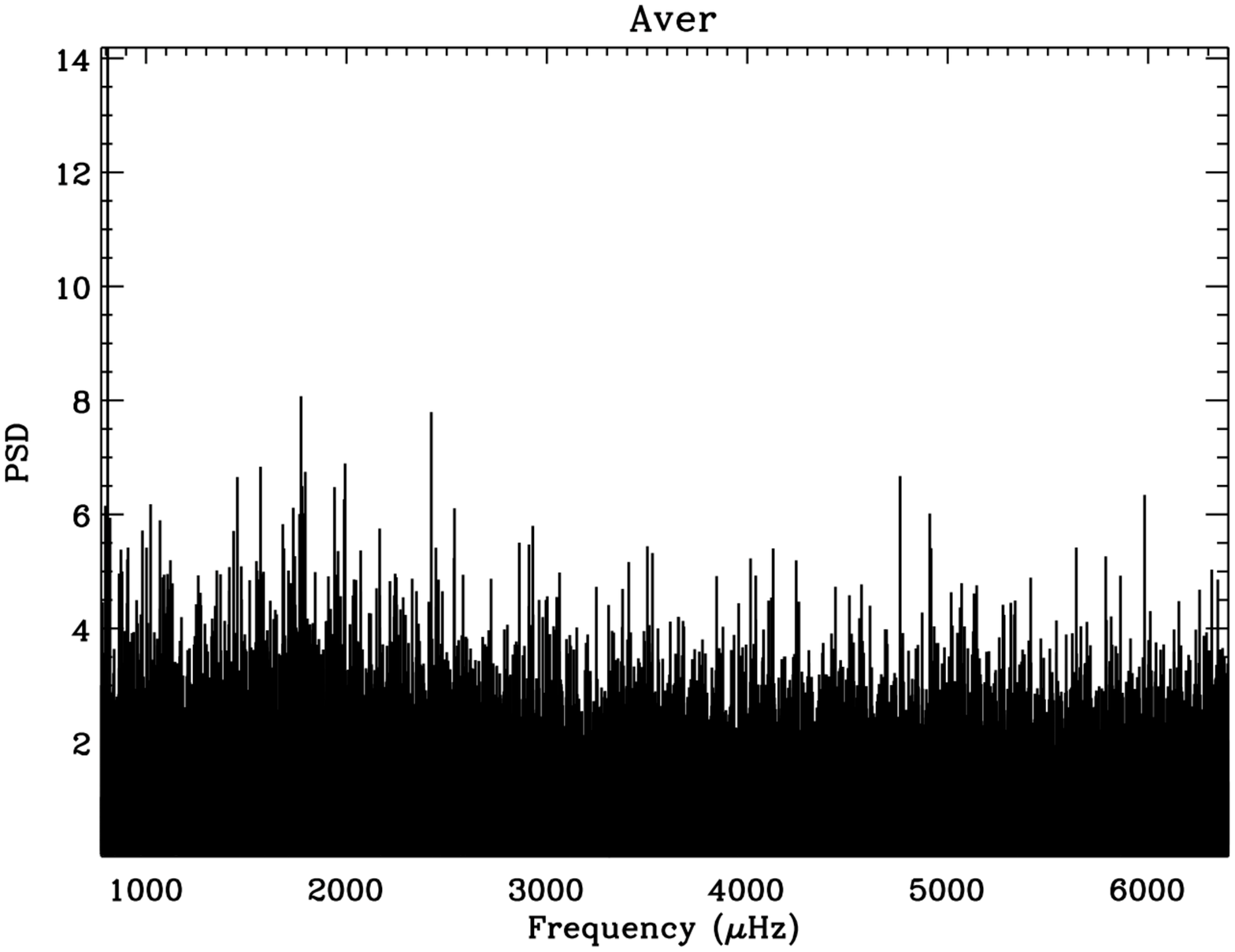}
\includegraphics[width=6.3cm, trim = 1cm 3.cm 2cm 9cm, angle=-90]{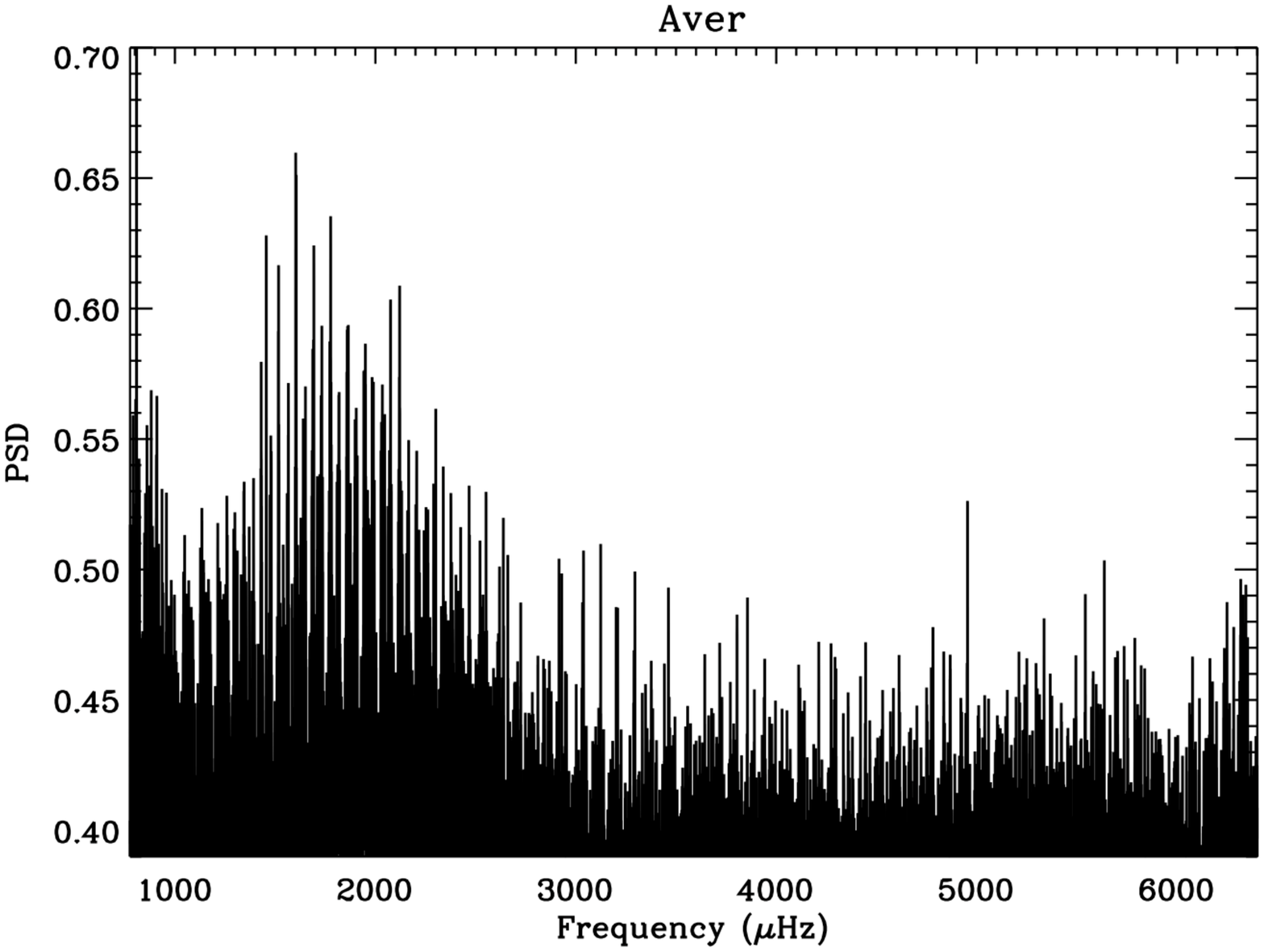}
\caption{Raw (top) and curvelet filtered (bottom) PSD of the low SNR target HD181906.}
\label{hd181906c}
\end{figure}

Fig.~\ref{hd181906c} shows the raw and the curvelet filtered PSD of HD181906 (Garc\'\i a et al. 2009). The reduction in the noise level of the filtered spectrum allows to retrieve the comb-like structure of the acoustic modes centered at ~1800 $\mu$Hz, with the p-mode hump visible from around 1000 to around  2600$\mu$Hz.

\begin{figure*}[!htbp]
\includegraphics[trim = 1cm 1cm 2cm 2cm, width=8.cm]{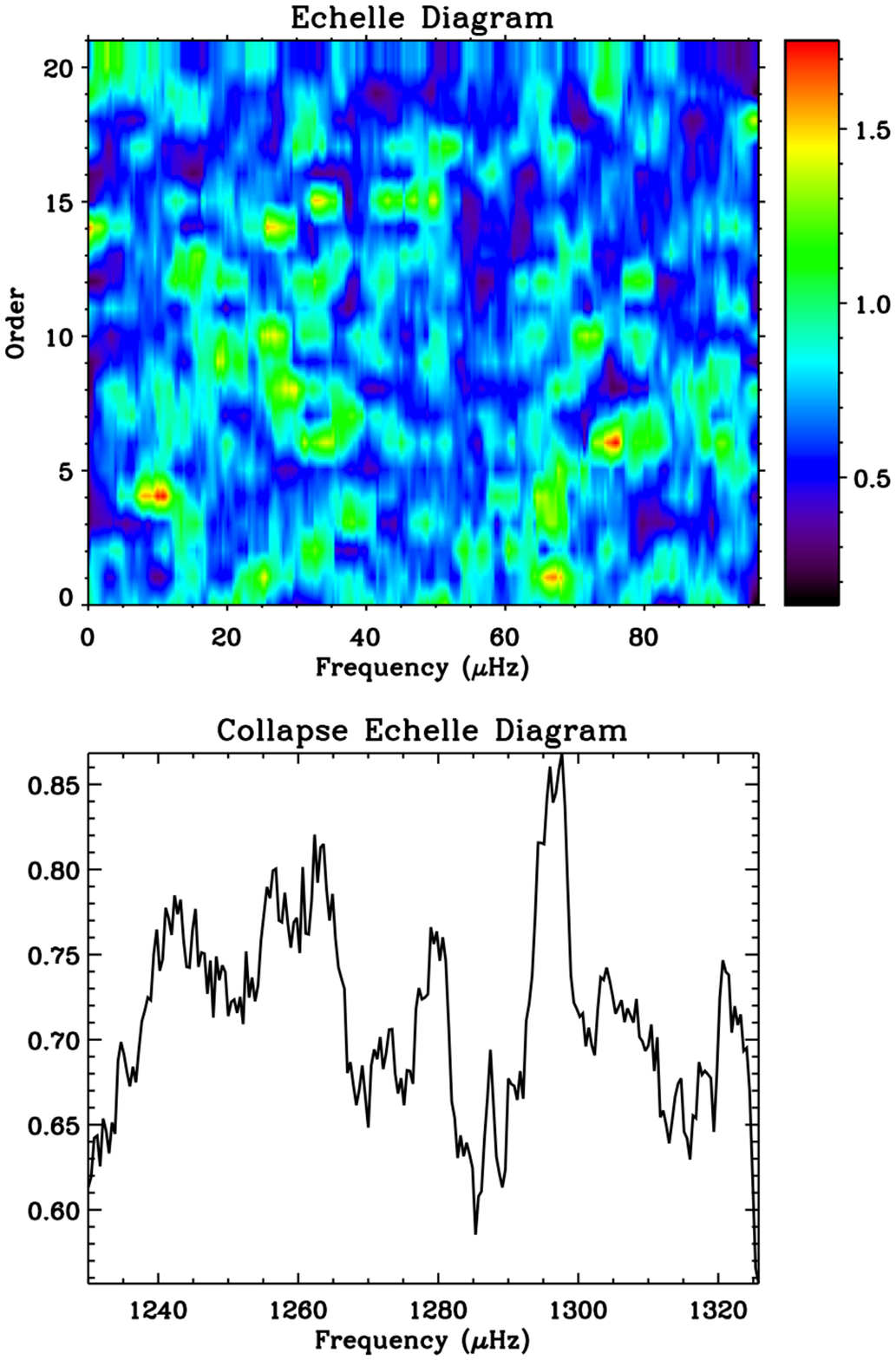}
\includegraphics[trim = 1cm 1cm 2cm 2cm, width=8.cm]{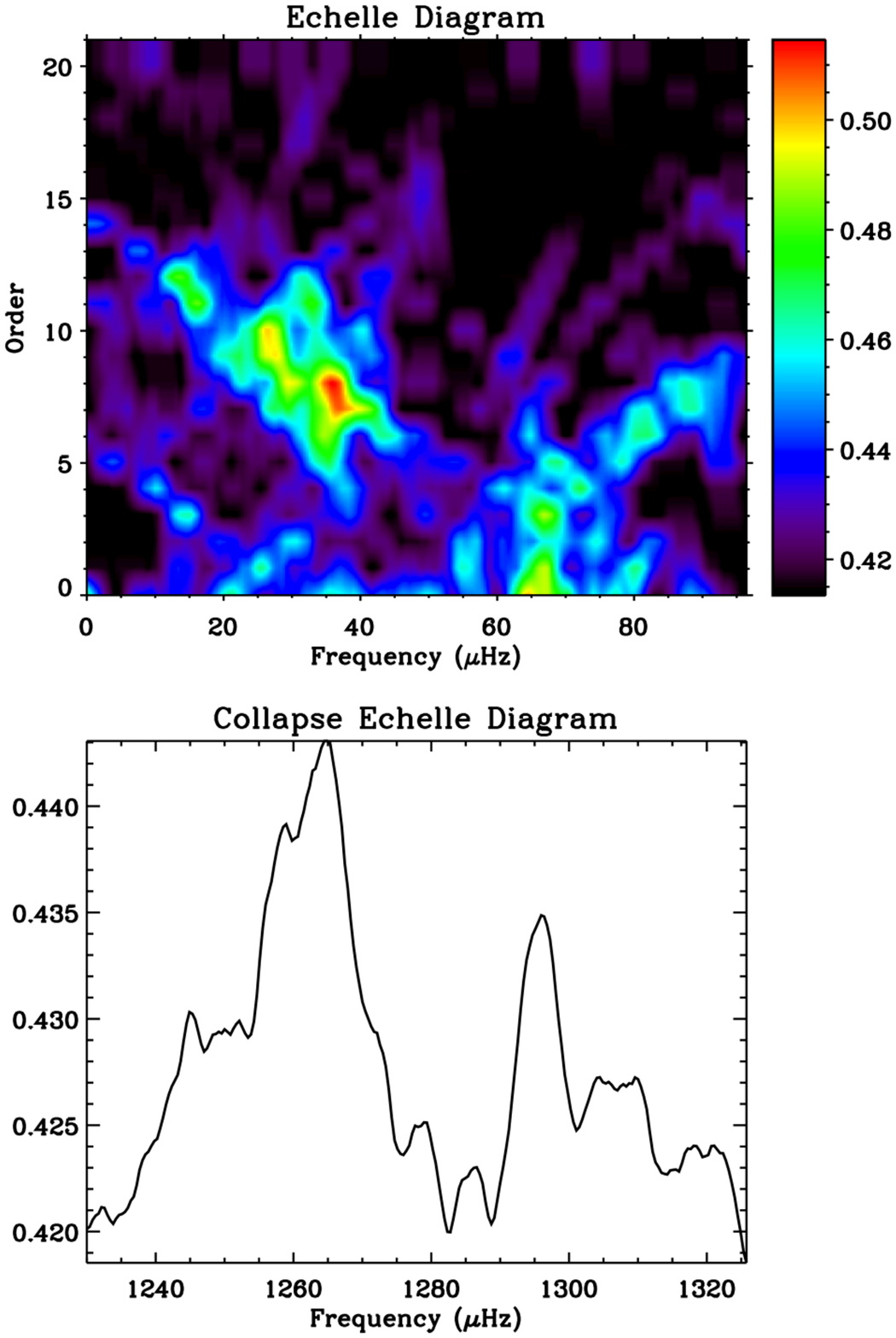}
\caption{Echelle diagram and collapsed spectrum computed as indicated in the text for the CoRoT short-run target HD175726: raw data (left panel) and curvelet filtered data (right panel)}
\label{figechelle}
\end{figure*}

The Curvelet filtering also works well for short runs, as in the case of HD175726 in which less than 30 days of observations were available (see Mosser et al. 2009). In Fig.~\ref{figechelle} we have plotted the original Echelle diagram and the collapsed spectrum (sum of each horizontal line of the Echelle diagram) for both, the raw data and the Curvelet filtered one. In the Curvelet filtered diagrams we can clearly identify the presence of two ridges corresponding to the odd and even modes.

\begin{figure*}[!htbp]
\includegraphics[width=13cm, trim = 1cm 3cm 2cm 15cm, angle=-90]{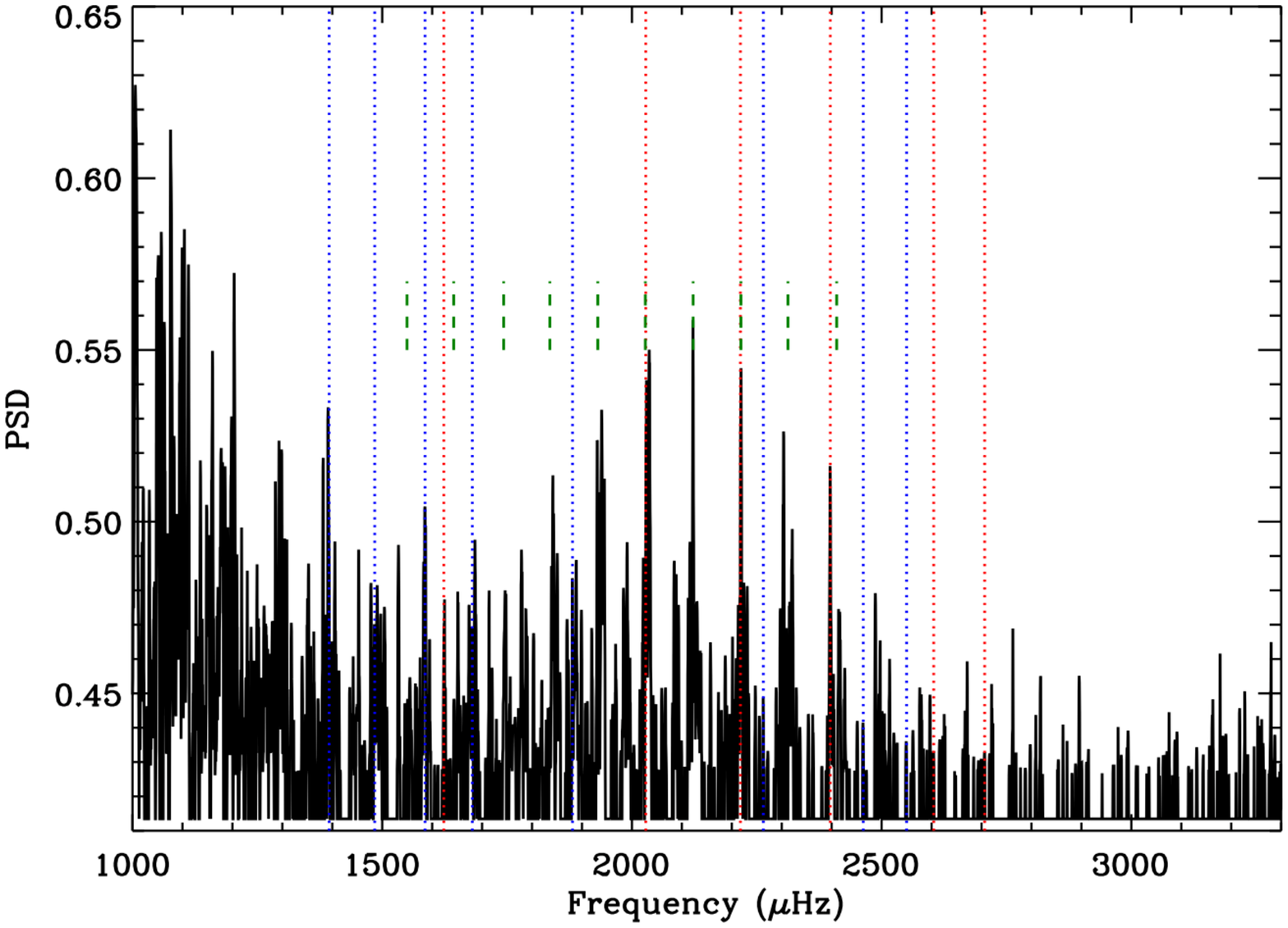}
\caption{Curvelet filtered PSD of HD175726. The dotted lines are the frequencies found by Mosser et al. (2009). Blue dotted lines are the frequencies $\nu_1$ in table 4 of Mosser et al. (2009), while the red ones are those labeled as $\nu_2$. Dashed green lines are the frequencies obtained by the partial reconstruction method described in section 2.}
\label{figMosser}
\end{figure*}

A comb-like pattern is clearly visible in the Curvelet filtered spectrum centred at~2120$\mu$Hz (see Fig.~\ref{figMosser}). We have compared the main peaks in the p-mode hump with those obtained by the partial reconstruction method described in section 2. The frequencies of the peaks obtained with both methods coincide very well, except for the two modes at low frequency for which the partial reconstruction method maintain a regular distance while in the Curvelet filtering spectrum the peaks are shifted from the regular spacing.

\section{Conclusions}
In this work we have shown two powerful methods to enhance the comb-like structure produced by acoustic modes propagating in the interior of stars. The partial reconstruction method allows to obtain the asymptotic large spacing by computing the PSPS and the reconstruction of the pattern in the PSD. The curvelet filtering technique uses the Echelle diagram to filter the noise and enhance the curved ridges of the odd and even modes. In both cases, the identification or tagging of the p-mode ridges is simplified and it can potentially help in the determination of the rotation speed and the inclination angle of the star, parameters that are correlated (Gizon \& Solanki 2003; Ballot, Garc\'\i a \& Lambert 2006) and are very difficult to disentangle even with modes observed with high SNR (e.g. Appourchaux et al. 2009; Deheuvels et al. 2010).  

Both methods have been intensively tested using AsteroFLAG simulated data (Chaplin et al. 2008) as well as on public CoRoT data, as shown here, and in the new low SNR target HD170987 (Mathur et al. 2010b) giving very promising results. Both techniques will be incorporated to our automatic asteroseismic pipeline (Mathur et al. 2010a) in order to analyze very low SNR targets of the survey phase of the Kepler mission (Bedding et al. 2010; Stello et al. 2010; Chaplin et al. 2010; Grigahcene et al. 2010). By doing so, we hope to extract a few individual frequencies in order to improve the determination of the stellar parameters in these low SNR targets (e.g. Stello et al. 2009).

\acknowledgements
CoRoT (Convection, Rotation and planetary Transits) is a minisatellite developed by the French Space agency CNES in collaboration with the Science Programs of ESA, Austria, Belgium, Brazil, Germany and Spain. This work has been partially funded by the GOLF/CNES grant at the CEA/Saclay.


\end{document}